\documentclass[conference]{IEEEtran}
\usepackage{url}  
\usepackage{graphicx}  
\usepackage{subfigure}
\usepackage{wrapfig}
\usepackage{algorithm}
\usepackage{algorithmic}
\usepackage{epstopdf}
\newtheorem{theorem}{Theorem}
\usepackage{amsmath}

\usepackage{fancyhdr} 
\usepackage{kantlipsum} 
\fancypagestyle{plain}{ 
	\fancyhf{} 
	\fancyhead[C]{Conference on \LaTeX} 
	 
} 
\usepackage{eso-pic} 

\ifCLASSINFOpdf
\else
\fi
\begin{document}
\AddToShipoutPictureBG*{ 
	\AtPageUpperLeft{ 
		\setlength\unitlength{1in} 
		\hspace*{\dimexpr0.5\paperwidth\relax}
		\makebox(0,-0.75)[c]{\textbf{Published in 2018 IEEE/ACM International Conference on Advances in Social Networks Analysis and Mining}}}} 
%
\title{Want to bring a community together? \\Create more sub-communities}

\author{\IEEEauthorblockN{Chen Luo}
\IEEEauthorblockA{
Rice University\\
Email: cl67@rice.edu}
\and
\IEEEauthorblockN{Anshumali Shrivastava}
\IEEEauthorblockA{
Rice University\\
Email: anshumali@rice.edu} 
}


%


\maketitle
\begin{abstract}
Understanding overlapping community structures is crucial for network analysis and prediction.  AGM (Affiliation Graph Model) is one of the favorite models for explaining the densely overlapped community structures. In this paper, we thoroughly re-investigate the assumptions made by the AGM model on real datasets. We find that the AGM model is not sufficient to explain several empirical behaviors observed in popular real-world networks. To our surprise, all our experimental results can be explained by a parameter-free hypothesis, leading to more straightforward modeling than AGM which has many parameters. Based on these findings, we propose a parameter-free Jaccard-based Affiliation Graph (JAG) model which models the probability of edge as a network specific constant times the Jaccard similarity between community sets associated with the individuals. Our modeling is significantly simpler than AGM, and it eliminates the need of associating a parameter, the probability value, with each community.  Furthermore, JAG model naturally explains why (and in fact when) overlapping communities are densely connected. Based on these observations, we propose a new community-driven friendship formation process, which mathematically recovers the JAG model. JAG is the first model that points towards a direct causal relationship between tight connections in the given community with the number of overlapping communities inside it. Thus, \emph{the most effective way to bring a community together is to form more sub-communities within it.}
The community detection algorithm based on our modeling demonstrates a significantly simple algorithm with state-of-the-art accuracy on six real-world network datasets compared to the existing link analysis based methods.
\end{abstract}


%
\IEEEpeerreviewmaketitle

\section{Introduction}
\label{sec:intro}

In a network structure, community means a set of groups of nodes that share common roles or interest, etc.
For example, in a protein-protein interaction network, each community in this network shows a common biological property.
In social networks, the community can be a family or a friendship circle, etc.
In the WWW (World-wide Web), web pages that have the same topic can belong to the same community.
Formally, a community is defined to be a set of cohesive nodes that have more connections inside than outside \cite{hoff2008modeling,fortunato2010community,kumar2017army}.

\begin{figure*}
	\centering
	\subfigure[Two Non-overlapped Communities]
	{\label{fig:no}\includegraphics[width=50mm]{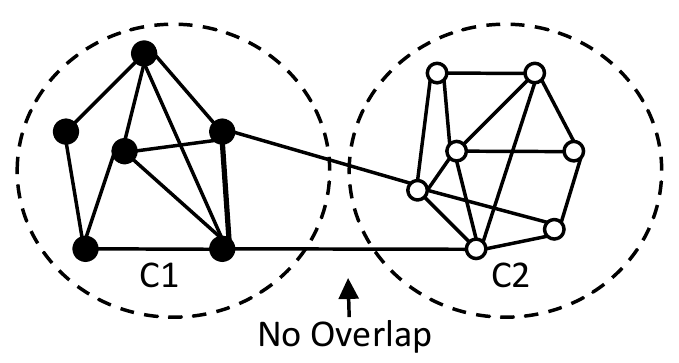}}
	\subfigure[Two overlapped Communities, with sparse link in overlap area]{\label{fig:os}\includegraphics[width=42mm]{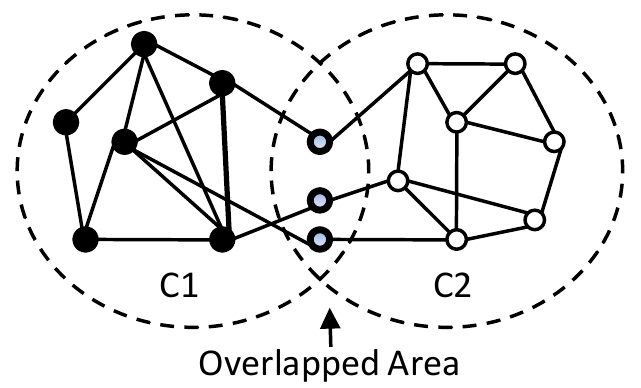}}
	\subfigure[Two overlapped Communities, with dense link in overlap area]{\label{fig:od}\includegraphics[width=42mm]{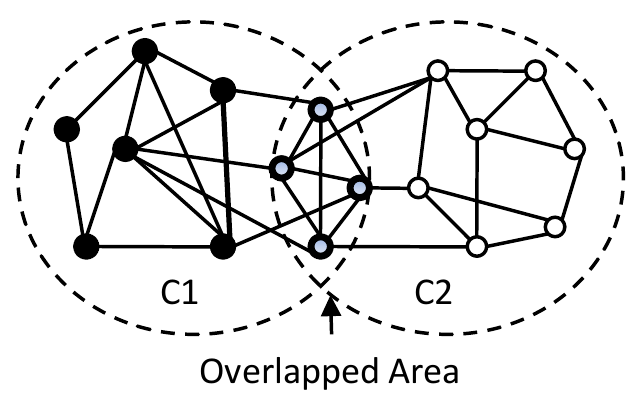}}
	\caption{Network community structures. In this paper, we focus on the community detection problem for densely overlapped structures.}\vspace{-0.2in}
	\label{fig:community}
\end{figure*}

Community detection algorithms are used for exploring the community structures in networks \cite{fortunato2010community}
, and are regarded as one of the most important tasks for network analysis \cite{farajtabar2016multistage,de2016learning,luo2014hetpathmine, luo2014semi}.
Community detection is of great interest in several real-world applications.
For example, the community structure can be used by recommender systems \cite{luo2014hete}.
The community structure of the protein-protein iteration networks help biologist better understanding the protein structure and properties \cite{rual2005towards}.
In social media analysis, the community structures of users can help us to understand user social behaviors\cite{agichtein2008finding}.

Prior network community detection methods such as Graph Partition \cite{fortunato2010community}, Modularity \cite{newman2006modularity}, etc., assume that in a graph, each node belongs to only one community. Fig \ref{fig:no} shows the non-overlap community structure in a network.
However, in most real-world networks, the communities are often overlapping \cite{xie2013overlapping}.
For example, in the online social media network Facebook. Each node means a person; each edge represents friendship relationship.
The community in a social network can be represented as different friendship circles.
It is evident that a person can belong to different friend circles (e.g., College Friends, and High School Friends) at the same time.
In a protein-protein network, proteins can have different biology properties, thus can simultaneously belong to several communities \cite{rual2005towards}.
As a result, understanding overlapping community structure of a graph is a crucial task for network analysis.

Because of the significance of the problem, many overlapping community detection methods have been proposed \cite{fortunato2010community,hoff2008modeling}.
Early attempts were heavily based on the assumption that the overlapping communities are less densely connected than non-overlap ones, as shown in Fig. \ref{fig:os}. However, with the availability of large real-data having ground truth labels, it was found that in most real-world social networks, the overlapping communities are more densely connected than non-overlapping ones \cite{yang2012community}.

Fig. \ref{fig:od} shows an example of the overlapping community structure where the overlapped part is densely connected.
There were strong empirical observations that probability of edges, in many real networks,  keeps increasing with increase in the number of shared community.
This observation motivates AGM model: Affiliation Graph Model for overlapping Network Community \cite{yang2012community}.

AGM assumes that individuals in each community $C_i$ have a \emph{community-specific} probability $p_i$ of forming connections. Furthermore, the probability is independent across different communities. This model naturally provided an explanation for the observation that overlapping communities are densely connected. The more the number of shared communities, the more the probability of the edge formation.  Based on AGM model, authors show an alternating sampling and optimization strategy for community detection, in a network. The model has one parameter per community. Thus, the number of parameters equal to the number of communities present in the network.  These parameters are optimized over data in alternating phase for community detection. AGM is currently one of the state-of-the-art algorithms in literature for community detection when only the edge information and no feature information about nodes is available.  AGM outperforms several other graph density based parameter free models.
 
In this paper, we point to a consistent observation (Section~\ref{sec:obs}), coming from six real-world networks, that cannot hold true under AGM modeling. Before we go into the actual empirical findings,  we illustrate our motivation by a simple example.

\subsection{Motivating Examples:}In a network $G(V,E)$, let $v_1. \ v_2, \ v_3 \in G$, be three individuals. $v_1$ belongs to the community set: $\{C_1, C_2, C_3, C_4, C_5\}$. $v_2$ belongs to community set: $\{C_1, C_6\}$, and $v_3$ belongs to the community set $\{C_1, C_7\}$.
Intuitively,  based on this information, we can expect that $v_2$ and $v_3$ are more likely to have an edge than $v_1$ and $v_2$ or $v_1$ and $v_3$.

However, according to AGM model \cite{yang2012community}, the edge probability of ($v_2$, $v_3$) and ($v_1$, $v_2$) (or $v_1$, $v_3$) should be exactly the same. This is because they all have only one shared community $C_1$ between any pair of them.  AGM model only considers the identity of shared community between the pair of individuals to decide on edge formation probability. Thus, with AGM, the edge formation probability between individuals has no dependence of what other involvements of individuals are.

Thus, $v_1$ in our example, even though involved in several other communities, will have the same chance of forming the friendship with $v_2$ as that of $v_3$. This behavior does not seem right, as $v_2$ and $v_3$ do not have other involvements and therefore are more likely to interact with each other compared to $v_3$.  In this paper, we show that indeed this is the case. We show empiric validation of this phenomenon, on 6 real-world networks, to demonstrate the above argument.

We conduct several validating experiments in section~\ref{sec:obs}. These experiments strongly suggest a nearly linear relationship between the edge probability and the Jaccard overlaps of the community sets of two individuals.
The same relationship holds at different granularity, and the patterns are consistent across domains.
From the empirical findings, it appears that the Jaccard similarity between the community sets of any two individuals is the ``sweet" measure, which in itself is sufficient to model the edge probability indicating a parameter-free model.

Strongly motivated by these empirical findings, we propose Parameter-free JAG (Jaccard-based Affiliation Graph) Model for modeling the community structures in social networks.
JAG model uses Jaccard similarity of the community sets associated with a pair of individuals as the probability of edge formation between them.
This modeling is significantly simpler than AGM, and it eliminates the need of associating a parameter, the probability value, with each community.
Furthermore, JAG naturally explains why (and in fact when) overlapping communities are densely connected. Individuals with significant overlap in their communities have high Jaccard Similarity and hence more edge formation probability.

We then show a simple random process of friendship formation based on community structure and random global preferences. Under fairly realistic assumptions, our theory recovers JAG model as the edge formation probability. To the best of our knowledge, this is the first theory that shows how preferences over communities lead to friendship formation. We also validate the consequences of these theoretical claims on all the six real datasets. The new theory and associated math could be of independent interest in itself.

\begin{figure}[t]
	\centering
	\subfigure[LiveJournal Network]{\label{fig:a}\includegraphics[width=39mm]{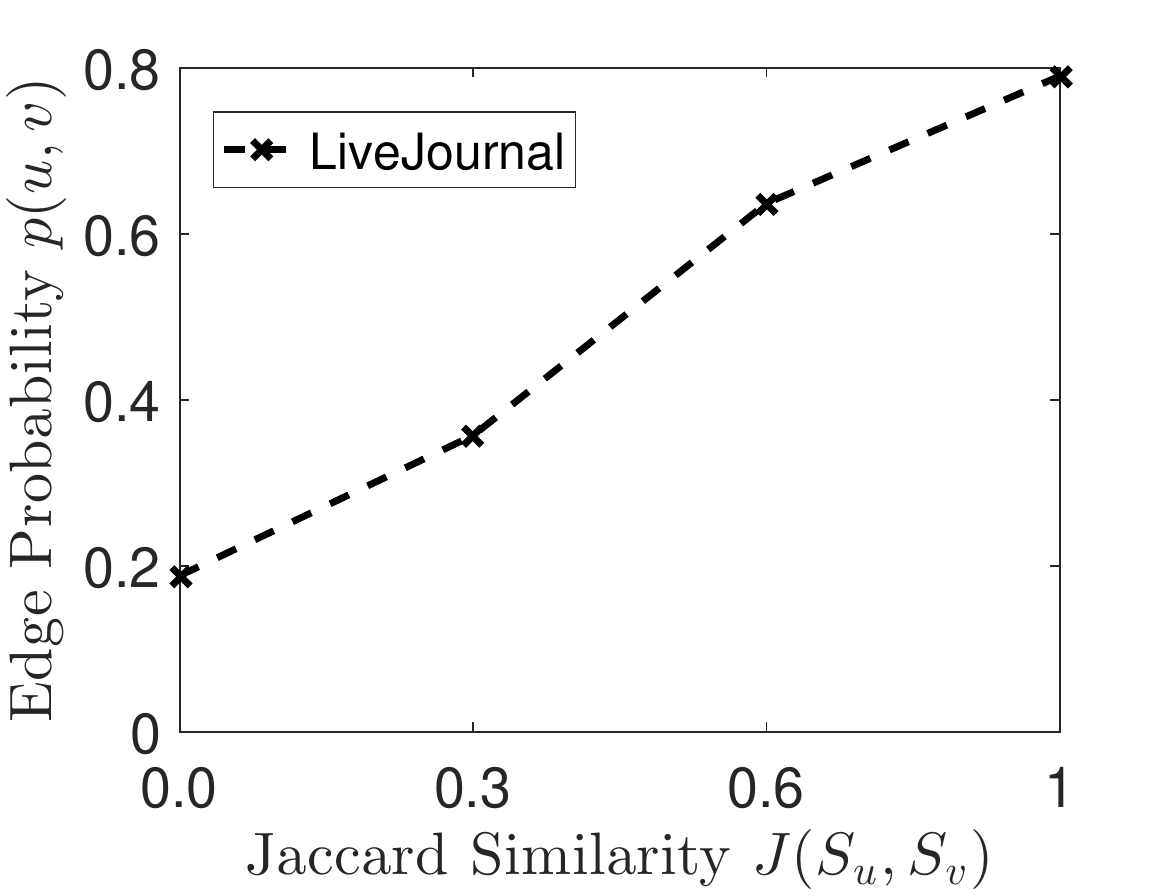}}
	\subfigure[Friendster Network]{\label{fig:b}\includegraphics[width=39mm]{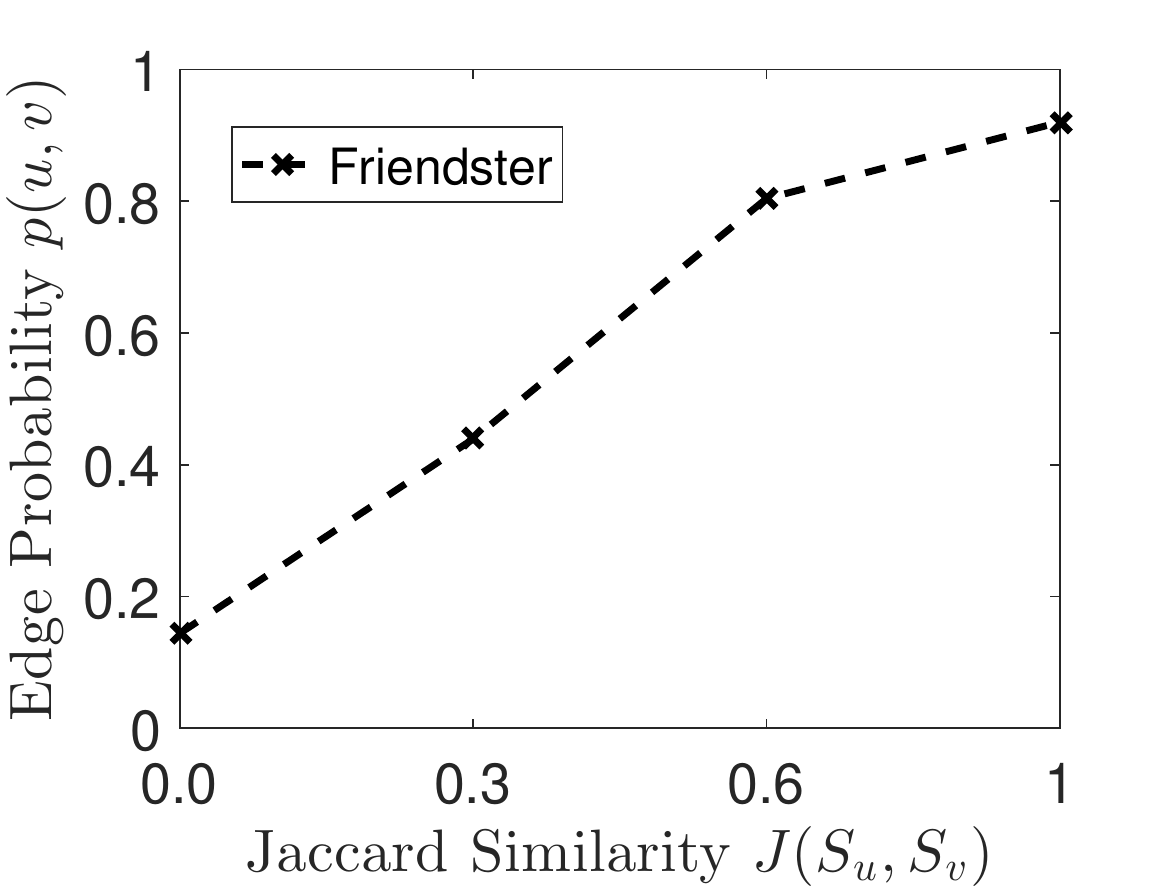}}
	\subfigure[Orkut Network]{\label{fig:b}\includegraphics[width=39mm]{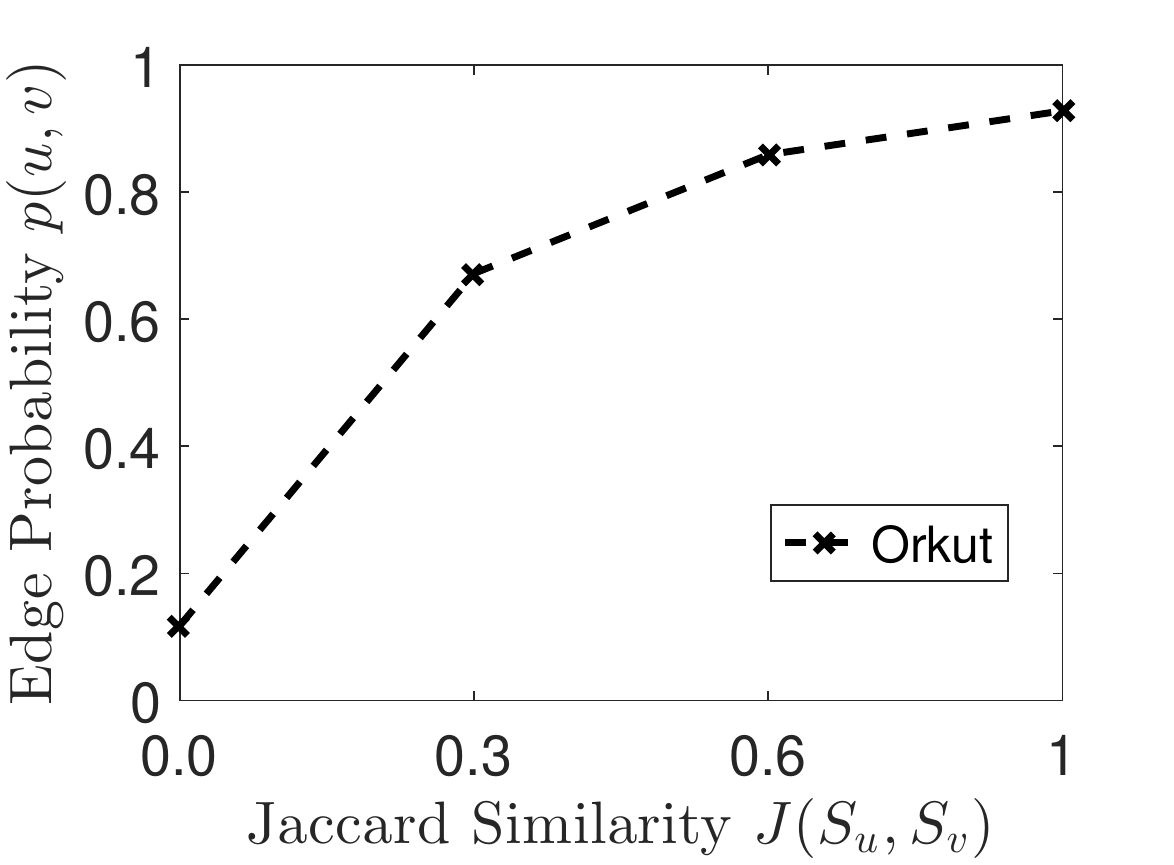}}
	\subfigure[Youtube Network]{\label{fig:b}\includegraphics[width=39mm]{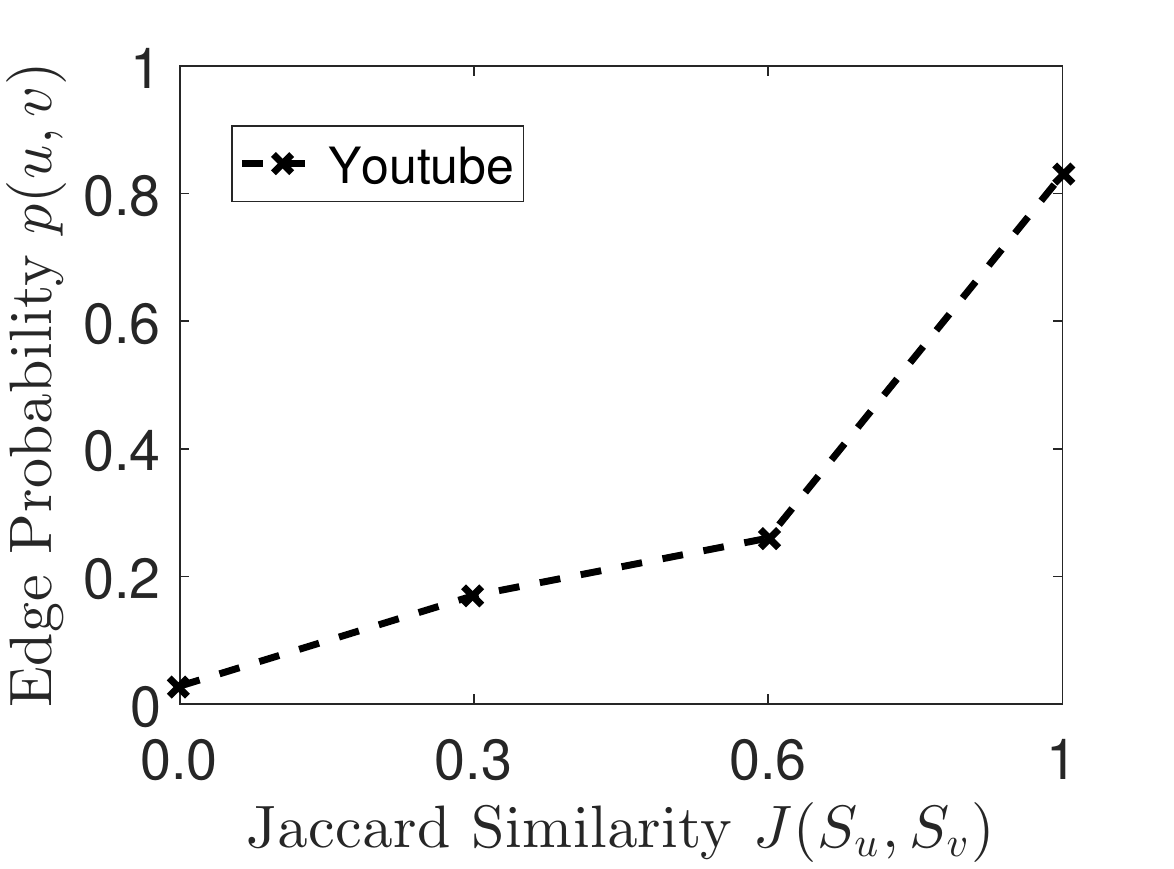}}
	\subfigure[DBLP Network]{\label{fig:b}\includegraphics[width=39mm]{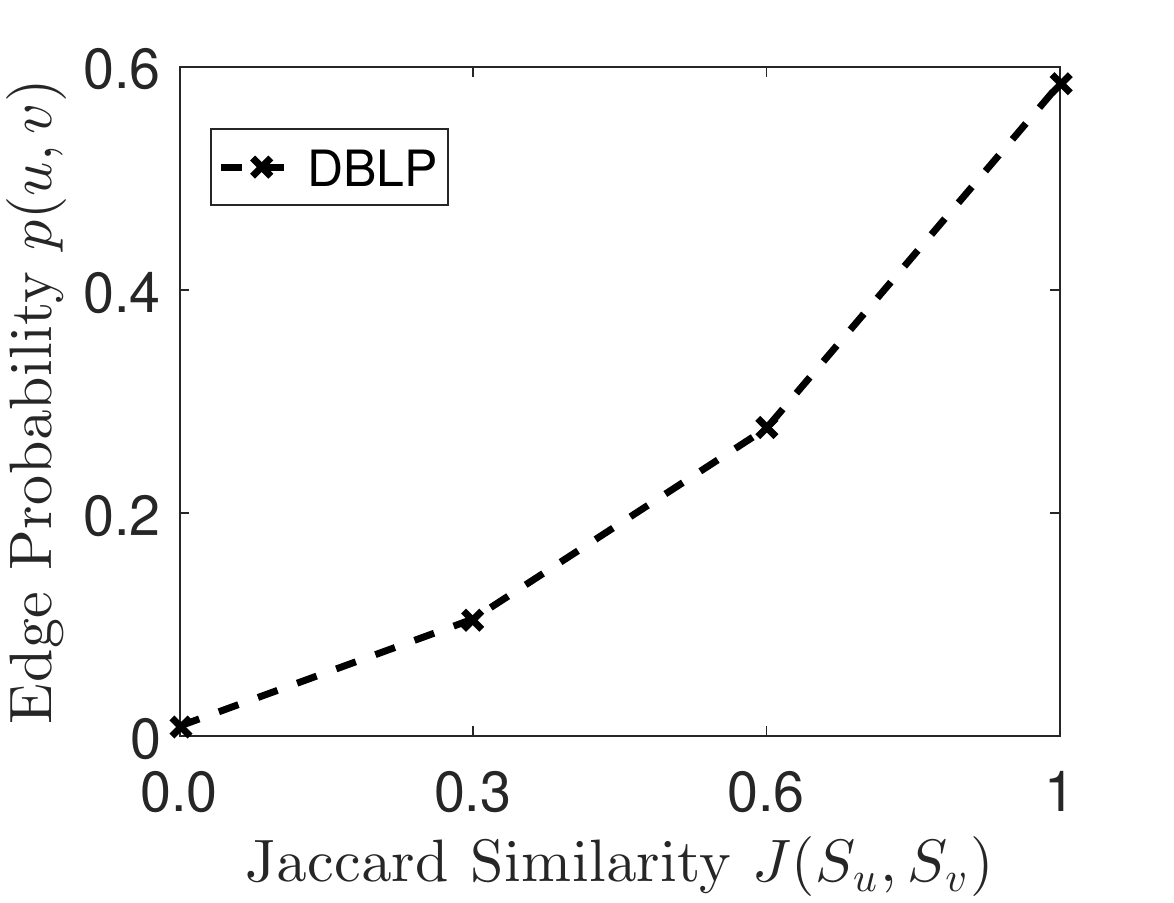}}
	\subfigure[Amazon Network]{\label{fig:b}\includegraphics[width=39mm]{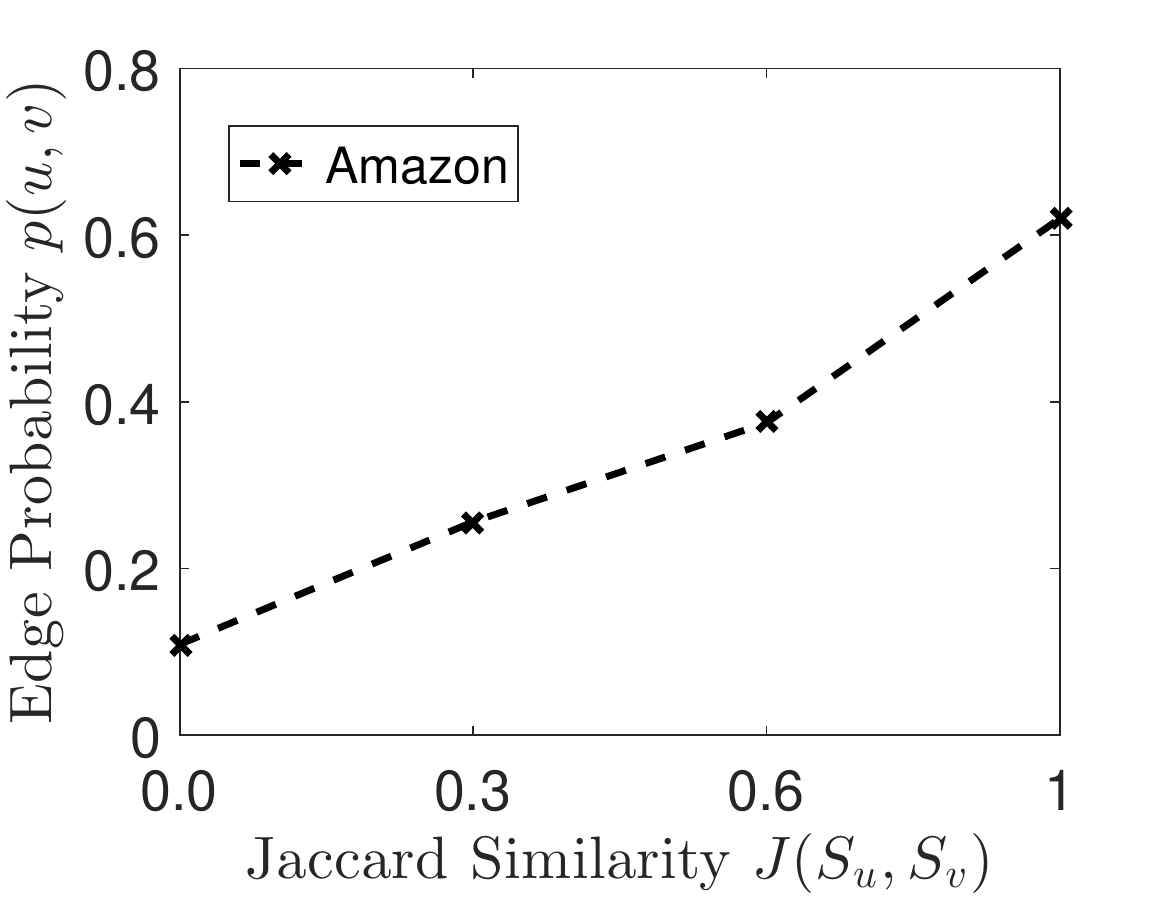}}
	\caption{We fix two large communities in each network dataset, and plot the edge formation probability $p(u,v)$ as a function of the Jaccard Similarity $J(S_u,S_v) = \frac{|S_u \cap S_v| }{|S_u \cup S_v| }$. Our process guarantees that the shared community of all the pairs is same. Thus, if AGM model was to hold true, then the probability of edge formation should be constant in all bins.  On the contrary, we observe a roughly linear relationship between the Jaccard similarity and the edge probability irrespective of the dataset used. Thus, we clearly see that AGM model is not good enough to explain this observation.}\vspace{-0.2in}
	\label{fig:local}
\end{figure}

Despite being able to explain more empirical findings, JAG modeling only requires one network specific global parameters compared to AGM which requires one parameter per community. As a consequence, JAG based community detection is significantly cheaper, and accurate, than the one based on AGM. We test JAG model on the task of community detection. JAG model, despite being cheap and parameter-free, consistently outperforms AGM on six different real datasets on three separate performance metrics.

There are a variety of community detection algorithms which use features and other side information for community detection~\cite{yang2013community}. However, our focus is on understanding communities only from the perspective of connectivity (and links) and not other side information. We will therefore only consider methodologies that exploit network connectivity for community detections, and we do not assume any other information.


\section{Background and Notations}
\label{sec:back}
Given an unweighted and undirected network $G(V,E)$, where $V = \{v_1,v_2,...,v_n\}$ is the set of nodes in $G$, and $E = \{e_1,e_2,...,e_m\}$ is the set of edges in $G$. $n=|V|$ is the number of node, and $m=|E|$ is the number of edges.

\subsection{AGM Model}
We firstly briefly review the AGM model~\cite{yang2012community}. AGM model assumes a simple generative process of edge formation based on coin flips. Given any two nodes $u$ and $v$ and let $C_{uv}$ be the set of shared communities between them. For every community $C_k \in C_{uv}$, we flip coins (independently) with community-specific probability $p_k$. The edge is formed if any of the coin flipped are in favor.

Under AGM modeling, given two nodes $u$ and $v$, the edge probability between $u$ and $v$ in AGM model is given by:
\begin{equation}
p(u,v) = 1-\prod_{k \in C_{uv}} (1 - p_k),
\end{equation}
where $C_{uv}$ is the communities common to both $u$ and $v$. $p_k$ is the edge formation probability of community $C_k$. The edge formation probability here denotes the probability that two nodes have an edge between them.

Clearly, from the definition, we can see that AGM assumes that each community $C_i$ had a community-specific probability $p_i$ of forming connections. Furthermore, the probability is independent across different communities. The number of parameters required by AGM modeling is therefore equal to the number of communities under consideration.

\subsection{Key Notations}
\label{sec:key}

\textbf{Community Set:} Given a node $u$, the community set $S_u$ is defined as the set of communities that node $u$ belongs to. For example, in a Facebook social network, the community set $S_u$ for a person $u$ is the set of his/her friendship circles: High-school Friend Community, Graduate-school Community, Family Community, etc.

\textbf{Jaccard Similarity}: The Jaccard similarity $J(S_u,S_v)$ between two community set $S_u$ and $S_v$ is defined as: \[ J(S_u,S_v) = \frac{|S_u \cap S_v|}{|S_u \cup S_v|},\] where $|S_u \cap S_v|$ is the number of shared nodes between $S_u$ and $S_v$, and $|S_u \cup S_v|$ is the number of all the nodes in the union set of $S_u$ and $S_v$.

\vspace{-0.05in}
\section{Observation Experiment}
\label{sec:obs}

\begin{table}
	\caption{The statistic of the six datasets.}
	\label{tab:datasets}
	\vskip -0.2in
	\begin{center}
		\begin{small}
			\begin{sc}
				\begin{tabular}{lcccr}
					\hline
					Data set & Node & Edge & Community \\
					\hline
					Live Journal    & 4.0 M & 34.9M & 310 K \\
					Friendster & 120 M & 2600 M & 1.5 M\\
					Orkut    & 3.1 M & 120 M & 8.5 M \\
					Youtube    & 1.1 M & 3.0 M &  30 K \\
					Amazon     & 0.34 M & 0.93 M & 49 K\\
					DBLP      & 0.43 M & 1.3 M & 2.5 K \\
					\hline
				\end{tabular}
			\end{sc}
		\end{small}
	\end{center}
	\vskip -0.1in
\end{table}

We use six publicly available real-world network datasets where we also have ground truth labels for communities of each node.
The datasets contain four online social networks: the LiveJournal blogging network, the Friendster online network, the Orkut social network, and the Youtube social network.
We also consider an Amazon product co-purchasing network, and the collaboration network: DBLP.
The basic statistic of the network datasets is shown in Table \ref{tab:datasets}.  It is pointed out that all the networks in the data sets are unweighted and undirected static graph.
The datasets are publicly available on \url{http://snap.stanford.edu}

\begin{figure}[t]
	\centering
	\subfigure[Live Journal Network]{\label{fig:a}\includegraphics[width=39mm]{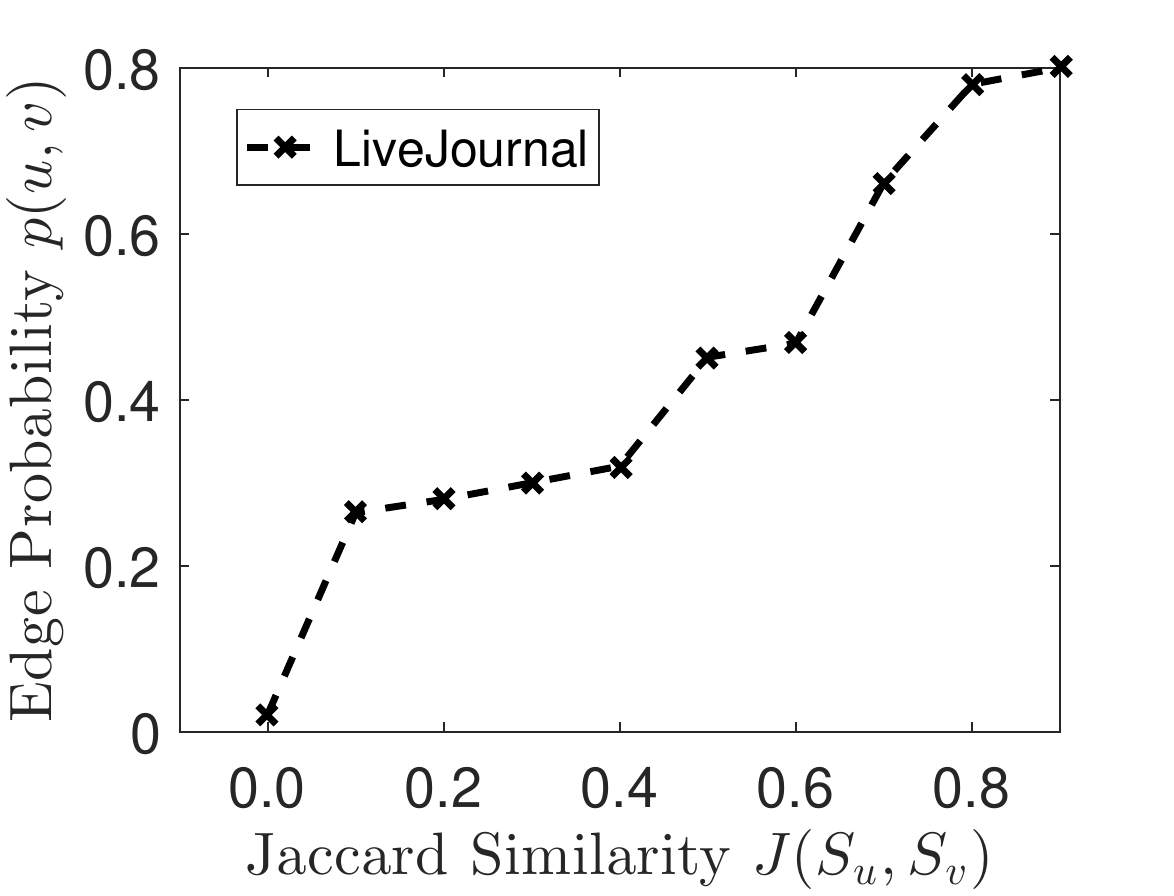}}
	\subfigure[Friendster Network]{\label{fig:b}\includegraphics[width=39mm]{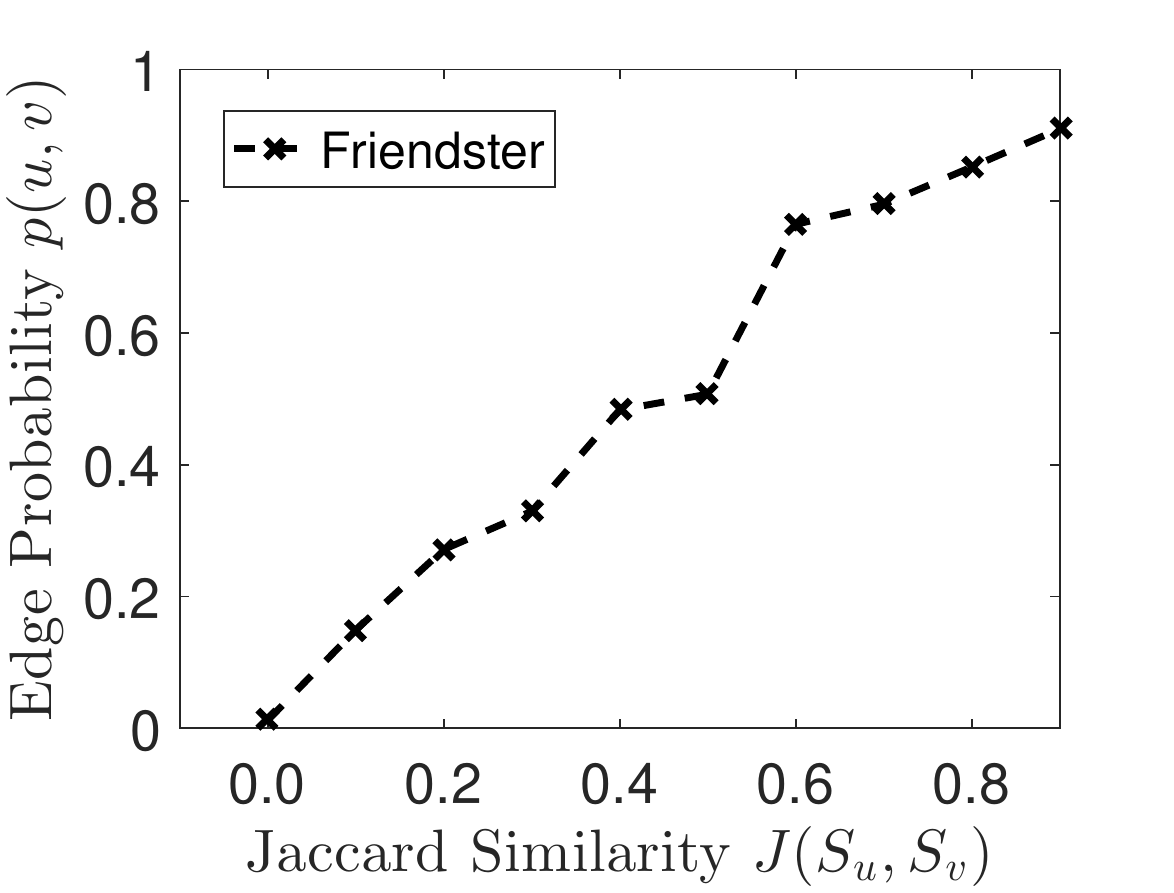}}
	\subfigure[Orkut Network]{\label{fig:b}\includegraphics[width=39mm]{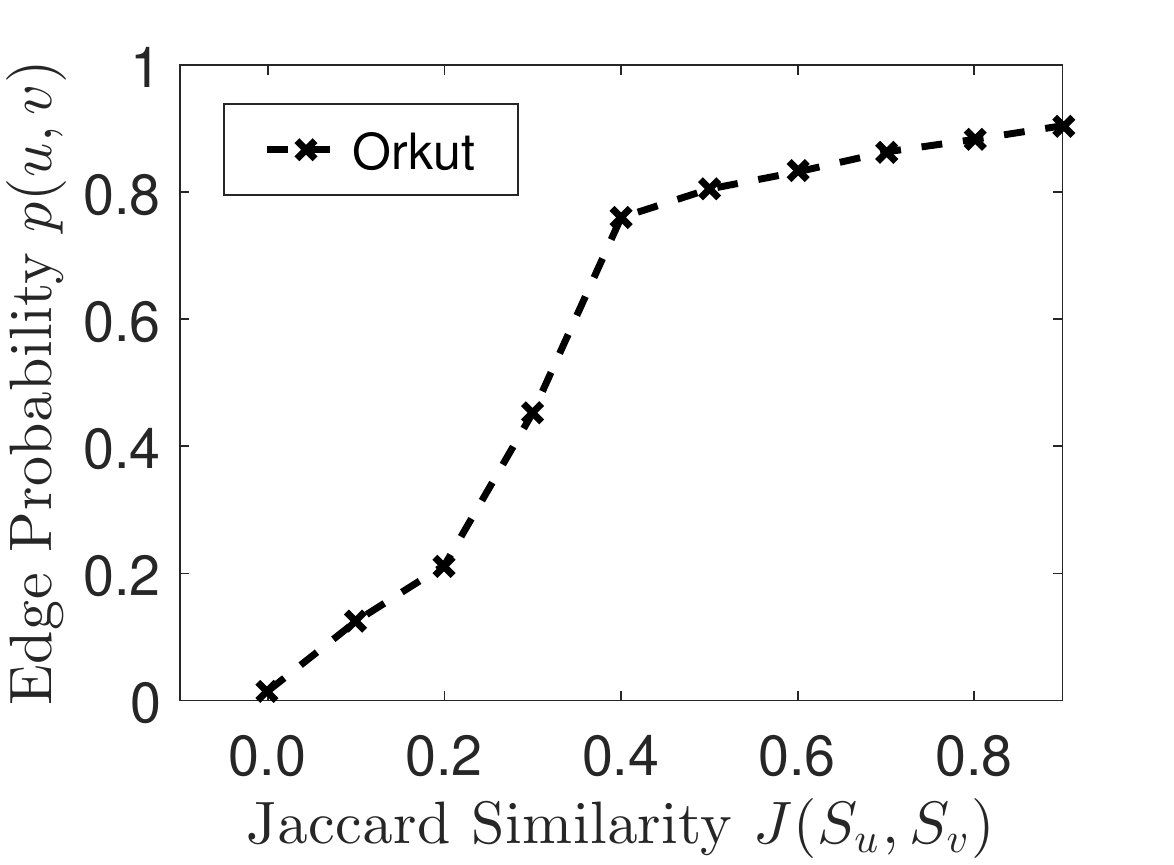}}
	\subfigure[Youtube Network]{\label{fig:b}\includegraphics[width=39mm]{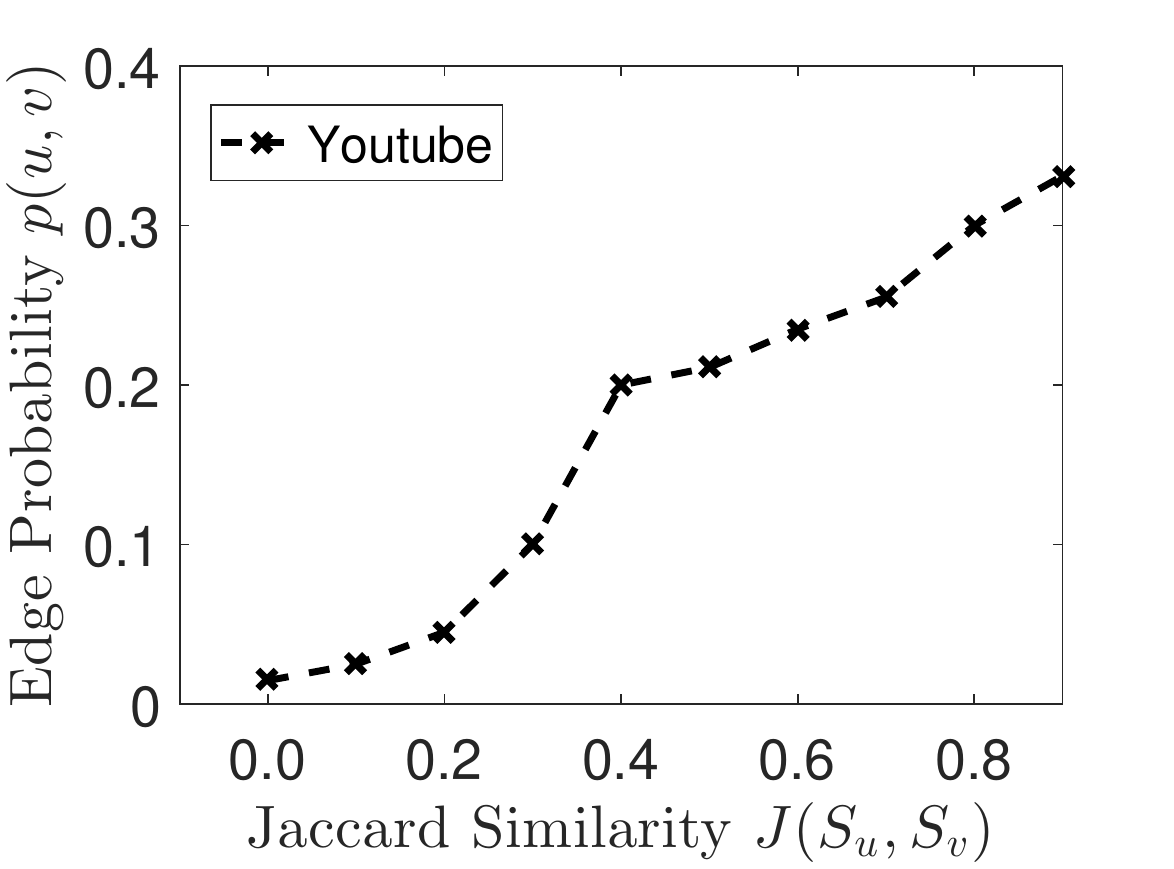}}
	\subfigure[DBLP Network]{\label{fig:b}\includegraphics[width=39mm]{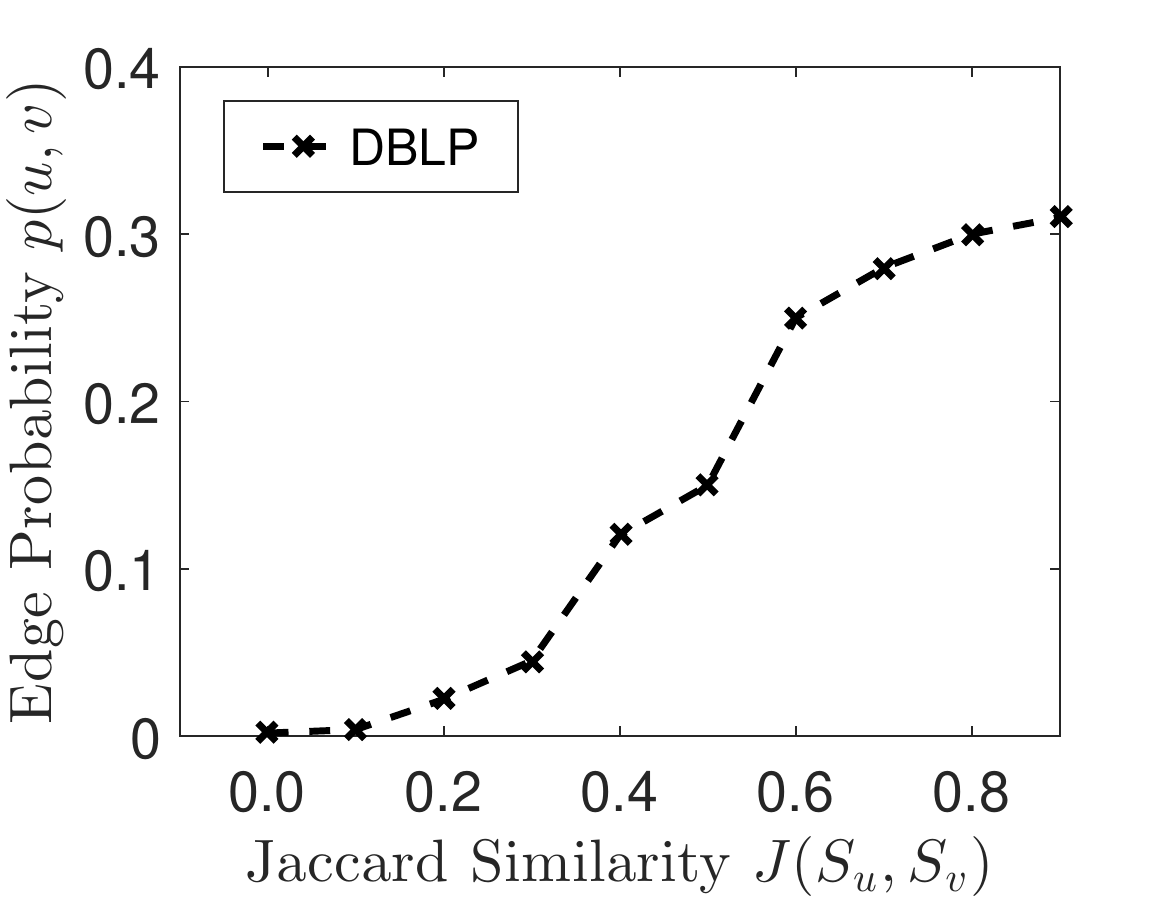}}
	\subfigure[Amazon Network]{\label{fig:b}\includegraphics[width=39mm]{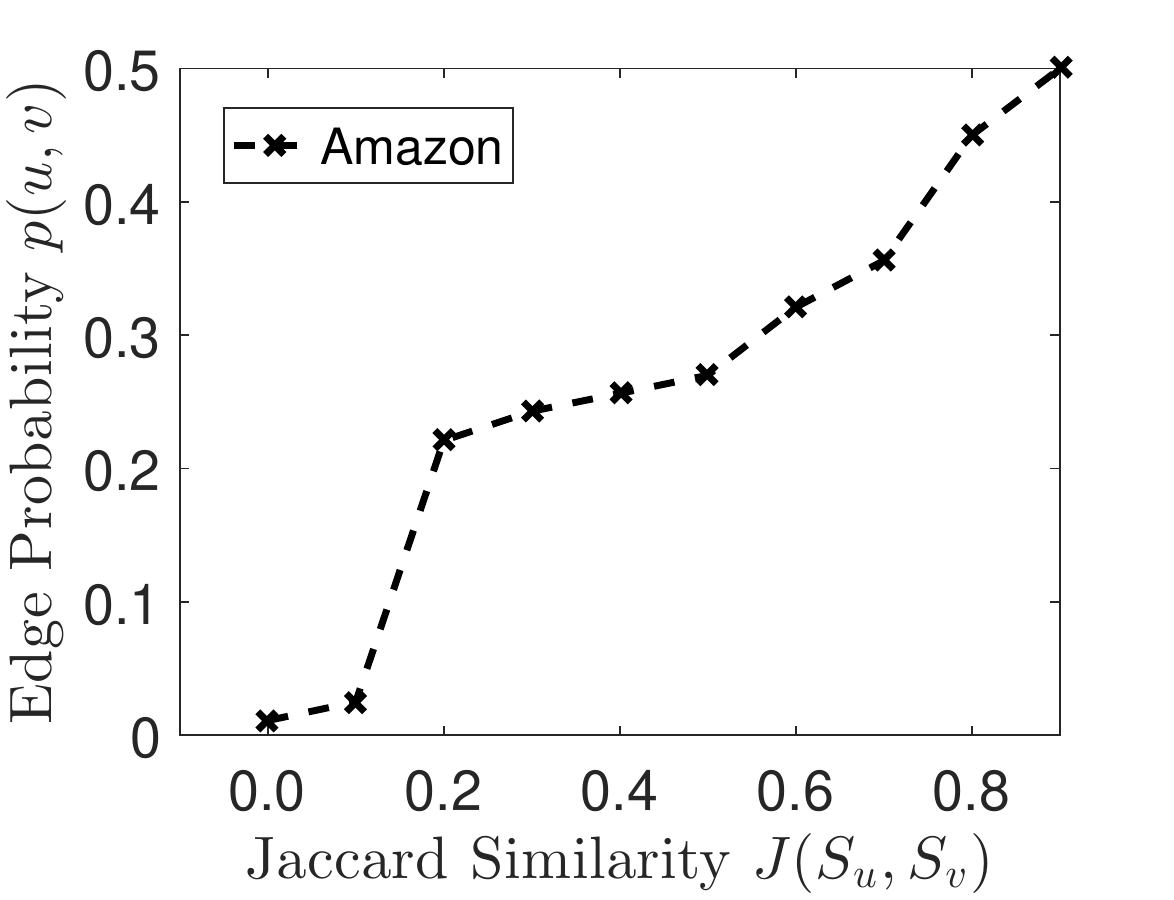}}
	\caption{We plot the edge formation probability $p(u,v)$ as a function of the Jaccard Similarity $J(S_u,S_v) = \frac{|S_u \cap S_v| }{|S_u \cup S_v| }$ on full networks. We can clearly see similar trends as shown in Fig.\ref{fig:local}. With the increasing of the Jaccard similarity, the probability keep increasing. In addition, we can also observe that the relationship between the Jaccard similarity and edge probability can be reasonably approximated by a linear relationship.}\vspace{-0.2in}
	\label{fig:global}
\end{figure}

We first provide the empirical motivation of our proposal. Based on our motivating example,  we hypothesize that the edge formation probability $p(u,v)$ between individuals $u$ and $v$ should be some unknown function of the Jaccard similarity between their community set, i.e.
$p(u,v) = f(J(S_u,S_v))$ instead of being just dependent on the number of shared communities. In other words, by such modeling, we ensure that the normalization matters. As an individual belonging to many communities will have less attention span on one of his community compared to an individual who only belongs to that community.

\subsection{Validation 1}
\label{sec:val1}

To validate this hypothesis we performed the following evaluation:
Given a network data, we first fix two large communities say $C_1$ and $C_2$ in them.  Then, we collect all the node pairs which only have these fixed communities in common. Note, the individuals $u$ and $v$ may belong to other communities, but the pair only shares the same \emph{fixed} two communities, i.e. $|S_u \cap S_v| = \{C_1, C_2\}$

We then plot the edge formation probability as a function of the Jaccard Similarity $J(S_u,S_v) = \frac{|S_u \cap S_v| }{|S_u \cup S_v| } = \frac{2}{|S_u \cup S_v| }$. To get such edge probabilities. We sample a pair of nodes satisfying
$|S_u \cap S_v| = \{C_1, C_2\}$ by rejection sampling~\cite{bishop2006pattern}.
Then for all those pairs, we bin them into groups based on their Jaccard similarity $J(S_u,S_v)$.  In all these bins we compute the probability of an edge, i.e. $p_{edge} = \frac{pairs \ with \ edge}{total \ pairs}$.
We ensure that we have sufficient statistics in each bin, by continuing to sample until we get large enough numbers.  We also used coarse-grained binning to ensure sufficient statistics.

We plot this probability $p_{edge}$ with Jaccard similarity for all the six datasets (as in Fig. \ref{fig:local}).  We see a roughly linear relationship between the Jaccard similarity and the edge probability irrespective of the dataset used.  This is very fascinating.

First of all, our process guarantees that the shared community of all the pairs is same. Thus, if AGM model was to hold true, then the probability of edge formation should be constant in all bins.  On the contrary, we observe a linear trend, i.e. the probability increases with Jaccard similarity of the associated community set. Thus, we clearly see that AGM model cannot explain these observations.

Based on the linear trend, we can hypothesize that our assumption of $p(u,v) = f(J(S_u,S_v))$  is in fact more like $p(u,v) = \alpha J(S_u,S_v))$, where $\alpha$ is a scalar which is network dependent. Note this modelling just has one parameter.

\subsection{Validation 2}
\label{sec:val2}

Our previous experiment used constrained pairs, where $|S_u \cap S_v| = \{C_1, C_2\}$. If our assumption was to be true, then we should observe similar behavior even for any pairs.  To see that, \emph{we repeated our experiment but this time sampling random pairs instead of forcing $|S_u \cap S_v| = \{C_1, C_2\}$.}  We again plot the probability of edge as a function of $J(S_u,S_v) $. This time we have more samples, and so, we can do fine grain binning of $J(S_u,S_v)$  values.

Fig. \ref{fig:global} shows the new plot again for all six data sets. We can clearly see similar trends.
With the increasing of the Jaccard similarity, the probability keeps increasing.  In addition, we can also observe from the Fig. \ref{fig:global} that, a linear relationship can reasonably approximate the relationship between the Jaccard similarity and edge probability.

\section{JAG: Parameter-free Jaccard-based Affiliation Graph Model}
\label{sec:jag}
Our observation experiment shows that given two nodes $u$, and $v$, and the corresponding community set: $S_u$ and $S_v$. The probability that node $u$ and $v$ have an edge is roughly linearly correlated with the Jaccard similarity between $S_u$ and $S_v$. We formalize this by proposing our edge probability as follow:
\begin{equation}
\label{equ:link}
p(u,v) = \alpha J(S_u,S_v).
\end{equation}
In above equation, $\alpha$ is a parameter that correspond to the properties of the network. We will discuss $\alpha$ later.

Coming back to our motivating example with  $v_1$, $v_2$ and $v_3$ as constructed in the first section.
With our model, the probability that $v_1$ and $v_2$ are connected is: $p(v_1,v_2) = \alpha/6$, and the probability that $v_1$ and $v_3$ are connected is $p(v_1,v_3) = \alpha/3$. Which means $v_2$ and $v_3$ are more likely to share an edge.  This follows our intuition that $v_2$ and $v_3$ are more likely to have common engagments leading to an edge compared to $v_1$ and $v_2$ or $v_1$ and $v_3$. Moreover, $p(u,v) = \alpha J(S_u,S_v)$ naturally explains why overlapping communities are generally dense.

\subsection{A New Simpler Theory of Friendship via Shared Communities}

Although our model is quite simple, it is not clear what process might lead to a linear relationship between community overlap (Jaccard similarity) and edge formation probability. In this section, we show a simple event based friendship formation process which mathematically leads to this linear relationship.

\begin{figure}[t]
	\centering
	\includegraphics[width=0.7\linewidth]{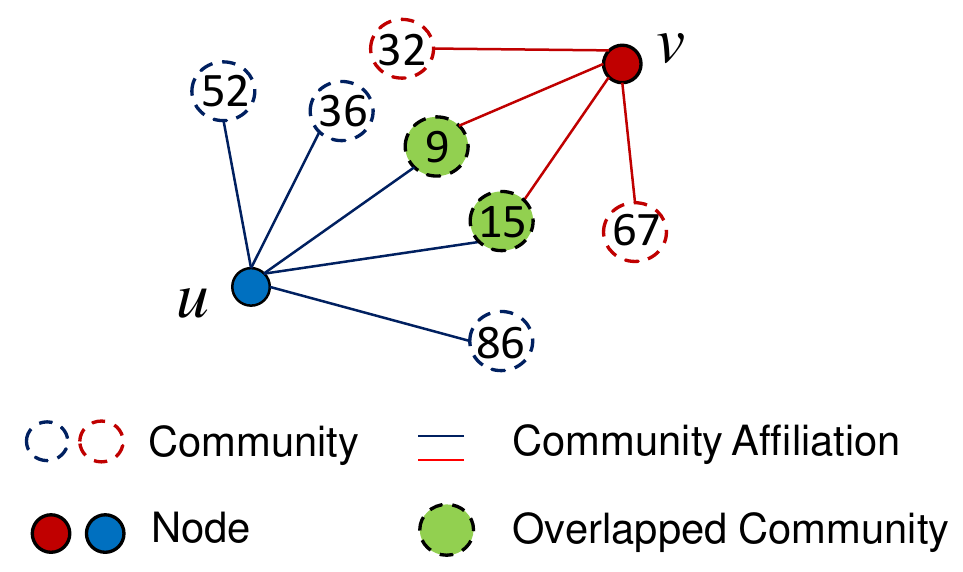}
	\caption{Example of edge formation process. Number inside each community denotes the corresponding preference ranking.}\vspace{-0.2in}
	\label{fig:validate}
\end{figure}

\textbf{The Friend Formation Process:} It is known that community events are a common place for people to meet and form friendships. We assume that activities keep happening. At a given time instance, a random ranking (total order) of preferences are assigned to the communities.  In particular, at a given instance nature samples a global ranking of communities in terms of preferences. This preference is uniformly chosen, among all possible preferences, at every instance. See Fig. \ref{fig:validate} for illustration.

Based on this sampled ranking, every individual $u$ selects the best-ranked community, from the set of the communities he is a member, i.e., $S_u$. Individual $u$ then chooses to participate in that preferred community event (or is attentive to the community). If two individuals $u$ and $v$ are in the same event, then there is a constant $Const$ chance (usually small) that they form a friendship.

\begin{theorem}
	Under the above friend formation process, the probability of edge between $u$ and $v$ is precisely a constant $\times$ the Jaccard Similarity between their community set $\frac{|S_u \cap S_v|}{|S_u \cup S_v|}$
\end{theorem}
\textbf{Proof:} For given individuals $u$ and $v$, consider the set of communities $S_u \cup S_v$ they both can participate. Both individuals will go to the same event, if and only if the most preferred event, out of $S_u \cup S_v$, belongs to the set  $S_u \cap S_v$. Note that the ranking is uniform and therefore the probability of $u$ and $v$ ending up in the same event is $\frac{|S_u \cap S_v|}{|S_u \cup S_v|}$.

The final constant $\alpha$ of our model is the $Const$  multiplied by the number of times that the sampling is performed. Note that the probability keeps increasing over time, which is consistent with observed social networks where the edges keep increasing.

\textbf{Global vs. Personal Preference:} It should be noted that our process assumes a global preference over the community events. This assumption is quite realistic on an average. There will be (local) variations, all the time due to personal preferences, but that will have negligible effects so long as a large number of individuals follow the average ranking.

\subsection{How do Connectivity Varies Across Communities?}
It is common to have some communities very tightly connected compared to others. AGM model assumes that every community has its own affiliation probability associated with it and edges are formed independently based on this probability. While AGM explains the existence of variations in the connectivity structure, our evidence suggests that it cannot explain the observed findings. Also, AGM requires more parameters, of the order of the number of communities,  to specify the connectivity probability.

Under AGM model, there can be two reasons for a given community $C$ to be densely connected: 1) The probability associated with the community is higher and 2) There are several overlapping communities inside $C$.

On the contrary, based on our model, which does not assume any community specific probability. The likelihood of forming an edge purely depends on the community overlaps. Under this simpler modeling, a community $C_1$ is tightly connected compared to another community $C_2$, if and only if, $C_1$ overlaps with more number of communities compared with $C_2$. In other words, $C_1$ has more sub-communities. Thus, our model implies a direct causal relationship between the existence of a larger set of overlapping communities and the density of the connections. This causation is also in line with our edge formation process driven by community preferences.  \emph{A direct causal relationship implies that an effective way of promoting friendship is to increase the community affiliation of every individual.}

If JAG model is correct, then different isolated communities, in a given network, should have same edge formation probability (or connectivity density). By isolated communities, we mean a community whose members are not part of any other community, or a community which does not overlap with any other community.  We next show this fact is true with surprising precision in all the six network datasets, providing yet another validation of our hypothesis.
\vspace{-0.1in}
\subsection{Validation 3}
\label{sec:val3}

Consider, \emph{Isolated Communities} which do not have any overlap with any other communities.  Then the Jaccard similarity between $S_u$ and $S_v$ of every node $u$ and $v$ in these isolated communities should be equal to $\frac{1}{1} =1$.  Thus the edge formation probability $p(u,v)$ for every node $u$ and $v$ in the isolated community, based on our model, should be
$
p(u,v) = \alpha J(S_u,S_v) = \alpha.
$
Note, $\alpha$ is also the slope of the observed plots in Figure~\ref{fig:global}.

If our model is correct, then we should observe this phenomenon empirically. We again use all the six real networks datasets with ground truth community labels. In all the six network datasets, we extracted the isolated communities. We randomly select 5 communities, where none of its members are part of any other community.

After extracting isolated communities, we calculated the edge probability inside each community. For each dataset, we report the mean and standard deviation of this probability over the extracted five isolated communities. Also, we report the slope of the lines from the Figures~\ref{fig:global}, which is also the estimated value of $\alpha$, on each of the six datasets. The results are shown in Table. \ref{tab:obs2}.

To our surprise, we can see that the standard deviations are quite small, which clearly indicates that, for a given network, isolated communities have the same edge formation probability. The same holds irrespective of the type of network. Furthermore, this edge formation probability is very close to the estimated slope, or $\alpha$, as suggested by Equation~\ref{equ:link}. These observations strongly validate all our claims.

Overall, we have a significantly simpler model, which explains several empirical findings (sections \ref{sec:val1}, \ref{sec:val2}, and \ref{sec:val3}) consistency on six different real datasets. To the best of our knowledge, there is no other known model which explains all our experimental observations.

\begin{table}[htp]
	\caption{Edge Formation Probability in Isolated Communities vs.  $\alpha$.}
	\label{tab:obs2}
	\begin{center}
		\begin{small}
			\begin{tabular}{l|c|c}
				\hline
				Data set & $\alpha$ & Edge Probability \\
				\hline
				Live Journal & 0.71 & $0.70 \pm 0.01$  \\
				Friendster & 0.74 & $0.75 \pm 0.02$ \\
				Orkut    & 0.85 & $0.84 \pm 0.01$ \\
				Youtube    & 0.26 & $0.26 \pm 0.02$ \\
				Amazon     & 0.32 & $0.31 \pm 0.03$\\
				DBLP      & 0.41 & $0.40 \pm 0.01$ \\
				\hline
			\end{tabular}
		\end{small}
	\end{center}
\end{table}
\section{From Model to Community Detection Algorithm}
\label{sec:alg}

\begin{figure}[t]
	\centering
	\subfigure[DELETING Operation]{\label{fig:delete}\includegraphics[width=90mm]{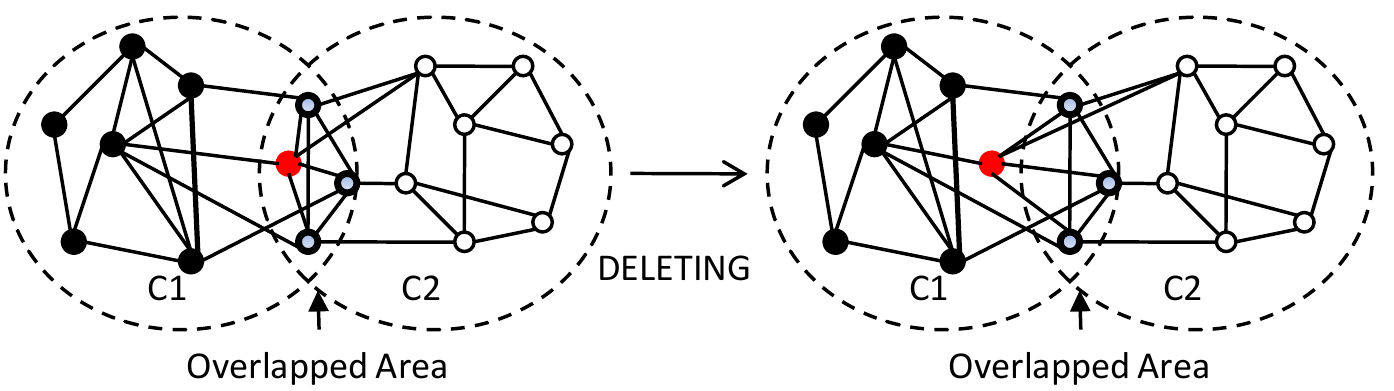}}
	\subfigure[ADDING Operation]{\label{fig:add}\includegraphics[width=90mm]{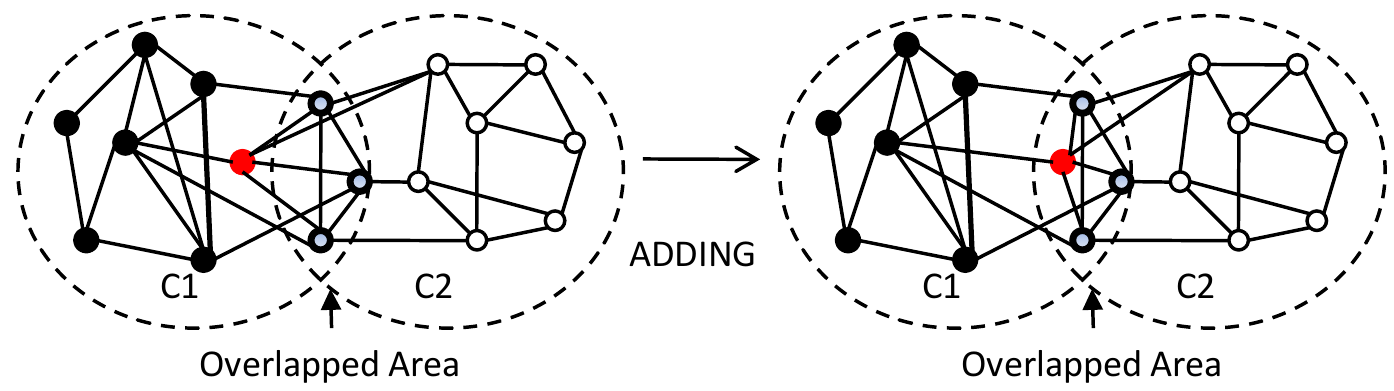}}
	\subfigure[SWITCHING Operation]{\label{fig:switch}\includegraphics[width=90mm]{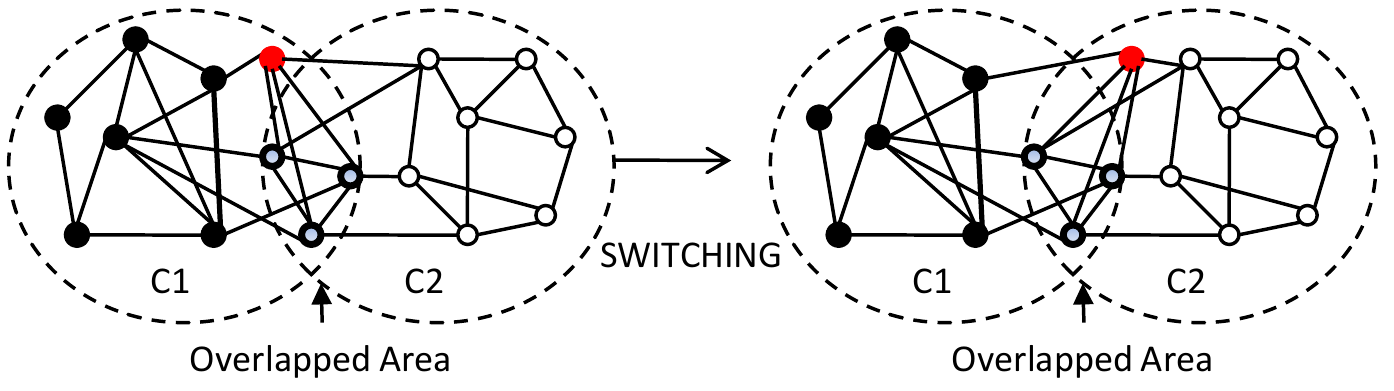}}
	\caption{Operations for the community assignment: DELETING, ADDING, and SWITCHING }\vspace{-0.2in}
	\label{fig:opr}
\end{figure}

\begin{algorithm}[t]
	\caption{\textbf{JAG Algorithm}}
	\label{alg:pfagm}
	\begin{algorithmic}
		\STATE {\bfseries Input:} $G(V,E)$, and $|C|$
		\STATE {\bfseries Output:} $A(V,C,M)$
		
		\STATE Initialize a random community assignment $A$
		\REPEAT
		\STATE Randomly choose an operation: {Deleting, Adding, and Switching.}
		\STATE Take the Randomly chosen operation on $A$ to generate $A'$.
		\STATE Calculate $\max_{\alpha} L(A)$ and $\max_{\alpha} L(A')$ use Equation. \ref{equ:loss}.
		\IF {$\max_{\alpha} L(A') > \max_{\alpha} L(A)$}
		\STATE Update $A = A'$
		\ELSE
		\STATE Update $A = A'$ with probability $\frac{\max_{\alpha} L(A')}{\max_{\alpha} L(A)}$.
		\ENDIF
		\UNTIL{ Convergence }
	\end{algorithmic}
\end{algorithm}

Having modeled the edge formation probability $p(u,v)$, given any community assignment $A(V,C,M)$, we can evaluate its likelihood.
The likelihood of any community assignment can be evaluated by the objective in Equation. \ref{equ:loss}.
\begin{equation}
\begin{split}
L(A) &= \prod_{(u,v)\in E} p(u,v) \prod_{(u,v) \notin E} (1-p(u,v))\\
&= \prod_{(u,v)\in E}\alpha J(S_u,S_v) \prod_{(u,v) \notin E} (1 - \alpha J(S_u,S_v))\\
&= \prod_{(u,v)\in E}\alpha \frac{|S_u \cap S_v|}{|S_u \cup S_v|} \prod_{(u,v) \notin E} (1 - \alpha \frac{|S_u \cap S_v|}{|S_u \cup S_v|})
\end{split}
\label{equ:loss}
\end{equation}
In the above, $L(A)$ denotes the likelihood of a community assignment $A(V,C,M)$. $S_u$ and $S_v$ can be obtained from the community assignment $A(V,C,M)$. As a result, we naturally obtain a community detection algorithm. The goal of community detection is to find the community assignment $A(V,C,M)$, and $\alpha$, which maximizes the likelihood.

\subsection{MCMC Based Assignment Generation}


Following \cite{yang2012community}, we use MCMC (Markov chain Monte Carlo)\cite{gilks1995markov} for updating community assignments.
Starting from a random assignment, we keep updating the community assignment $A(V,C,M)$ until we cannot improve the likelihood anymore.

To update the assignment $A$,  we rely on the set of `OPERATIONS' as defined in \cite{yang2012community}. For completeness we reiterate them here.
There are three types of operation for generating a new assignment $A'$. The three operations are showed in Fig. \ref{fig:opr}. The three operations are DELETING, ADDING, and SWITCHING:

\begin{itemize}
	\item \textbf{DELETING} operation is randomly select an node community pair $(v,c)$ in $M$, and delete the edge from $M$. In other words, randomly select a non-zero element in node community relation matrix $M'$, and set this entry as $0$; Fig. \ref{fig:delete} illustrate the deleting process.
	\item \textbf{ADDING} operation is randomly choose an node community pair $(v,c) \notin M$, and add it to $M$. In other words, randomly select a zero element in node community relation matrix $M'$, and set this entry as $1$. \ref{fig:add} illustrate the adding process.
	\item \textbf{SWITCHING} operation is randomly select two pairs $(v,c_1) \in M$ and $(v,c_2) \notin M$. Then we delete this the pair $(v,c_1)$, and add the pair $(v,c_2) \notin M'$. \ref{fig:switch} illustrate the switching process.
\end{itemize}

After generating a new proposal assignment $A'(V,C,M')$, we then calculate the likelihood value of $\max_{\alpha} L(A')$. If $\max_{\alpha} L(A')$ is greater than $\max_{\alpha} L(A)$, we accept this new community assignment, else, we accept this community assignment with the probability $\frac{\max_{\alpha} L(A')}{\max_{\alpha} L(A)}$. This leads to our proposed algorithm: JAG Algorithm.

\subsection{JAG Algorithm}

The overall JAG algorithm is summarized as in Algorithm \ref{alg:pfagm}. The input of the JAG algorithm contains the unlabeled network structure $G(V, E)$, the number of communities $|C|$. Since there is only one parameter $\alpha$, we have two options: 1) We can alternate between MCMC and optimizing $\alpha$, just like~\cite{yang2012community} or 2) We can simply use a grid search to find the $\alpha$ in $[0,1]$, that maximizes the likelihood. We use the later, as it is quite easy. Furthermore, there are no concerns about non-convexity of the objective.

AGM, on the other hand, needs to alternate between an MCMC and a costly non-convex vector optimizations which requires another relaxation leading to further approximations to solve it. See~\cite{yang2012community} for details.

\subsection{Implementation Details}
\label{sec:prac}

\textbf{Local Minimum}
JAG algorithm is a Markov chain searching processing. It is easy to get to a local minimum. So that in practice, we choose a number of initial community assignment and do the Markov chain searching. In addition, at each Markov chain search step, we can usually generate a number of community assignment, and accept the best community assignment. The final result is the community assignment that has the maximum value of Equation .\ref{equ:loss}.

\textbf{Convergence}
Although there no rigorous theoretical guarantees for the convergence of our model. But in our experiment, the Markov chain searching process can quickly convergences (A few seconds for a thousand level sampled small graph.). In addition, even for the more complexity model in \cite{yang2012community}, they also show quickly convergence in practice.

\begin{table*}[t]
	
	\begin{minipage}{0.33\linewidth}
		\centering
		\caption{Live Journal Network}
		\tabcolsep=0.12cm
		\begin{tabular}{lcccr}
			\hline
			Methods & F1-Score & NMI & Omega Index \\
			\hline
			JAG    & \textbf{.757$\pm$ .01}& \textbf{.612$\pm$ .02}& .\textbf{723$\pm$ .03} \\
			AGM & .751$\pm$ .02& .565$\pm$ .01& .637$\pm$ .02 \\
			
			LC    & .455$\pm$ .02& .452$\pm$ .02& .698$\pm$ .01 \\
			CPM    & .557$\pm$ .01& .511$\pm$ .02& .658$\pm$ .02 \\
			MMSB   & .712$\pm$ .02& .315$\pm$ .02& .719$\pm$ .02 \\
			\hline
		\end{tabular}
	\end{minipage}
	\begin{minipage}{0.33\linewidth}
		\centering
		\caption{Friendster Network}
		\tabcolsep=0.12cm
		\begin{tabular}{lcccr}
			\hline
			Methods & F1-Score & NMI & Omega Index \\
			\hline
			JAG    & \textbf{.756$\pm$ .01}& \textbf{.725$\pm$ .02}& \textbf{.716$\pm$ .01}& \\
			AGM & .742$\pm$ .06& .574$\pm$ .02& .647$\pm$ .02 \\
			
			LC    & .348$\pm$ .02& .482$\pm$ .02& .542$\pm$ .02 \\
			CPM    & .327$\pm$ .02& .532$\pm$ .02& .681$\pm$ .03 \\
			MMSB   & .612$\pm$ .01& .315$\pm$ .01& .426$\pm$ .02 \\
			\hline
		\end{tabular}
	\end{minipage}
	\begin{minipage}{0.33\linewidth}
		\centering
		\caption{Orkul Network}
		\tabcolsep=0.12cm
		\begin{tabular}{lcccr}
			\hline
			Methods & F1-Score & NMI & Omega Index \\
			\hline
			JAG    & .714$\pm$ .01& \textbf{.635$\pm$ .02}& \textbf{.759$\pm$ .01} \\
			AGM & .734$\pm$ .01& .577$\pm$ .02& .658$\pm$ .03 \\
			
			LC    & .449$\pm$ .03& .484$\pm$ .06& .547$\pm$ .01 \\
			CPM    & .512$\pm$ .01& .498$\pm$ .01& .612$\pm$ .03 \\
			MMSB   & \textbf{.742$\pm$ .01}& .365$\pm$ .02& .516$\pm$ .02 \\
			\hline
		\end{tabular}
	\end{minipage}
	\begin{minipage}{0.33\linewidth}
		\centering
		\caption{Youtube Network}
		\tabcolsep=0.12cm
		\begin{tabular}{lcccr}
			\hline
			Methods & F1-Score & NMI & Omega Index \\
			\hline
			JAG    & .704$\pm$ .01& \textbf{.498$\pm$ .02}& \textbf{.709$\pm$ .01} \\
			AGM & .687$\pm$ .03& .425$\pm$ .02& .636$\pm$ .02\\
			
			LC    & .265$\pm$ .02& .482$\pm$ .01& .625$\pm$ .01 \\
			CPM    & .162$\pm$ .02& .181$\pm$ .01& .576$\pm$ .02 \\
			MMSB   & \textbf{.721$\pm$ .03}& .215$\pm$ .02& .619$\pm$ .03 \\
			\hline
		\end{tabular}
	\end{minipage}
	\begin{minipage}{0.33\linewidth}
		\centering
		\caption{DBLP Network}
		\tabcolsep=0.12cm
		\begin{tabular}{lcccr}
			\hline
			Methods & F1-Score & NMI & Omega Index \\
			\hline
			JAG    & \textbf{.761$\pm$ .02}& \textbf{.528$\pm$ .01}& \textbf{.702$\pm$ .01} \\
			AGM & .707$\pm$ .02& .408$\pm$ .06& .652$\pm$ .01 \\
			
			LC    & .458$\pm$ .03& .182$\pm$ .02& .521$\pm$ .04 \\
			CPM    & .428$\pm$ .02& .539$\pm$ .02& .646$\pm$ .01 \\
			MMSB   & .752$\pm$ .06& .510$\pm$ .02& .583$\pm$ .03 \\
			\hline
		\end{tabular}
	\end{minipage}
	\begin{minipage}{0.33\linewidth}
		\centering
		\caption{Amazon Co-Purchasing Network}
		\tabcolsep=0.12cm
		\begin{tabular}{lcccr}
			\hline
			Methods & F1-Score & NMI & Omega Index \\
			\hline
			JAG    & \textbf{.648$\pm$ .02} & \textbf{.575$\pm$ .02}& .572$\pm$ .01 \\
			AGM & .617$\pm$ .06& .526$\pm$ .04&  .495$\pm$ .01 \\
			
			LC    & .556$\pm$ .02& .568$\pm$ .02& .201$\pm$ .01 \\
			CPM    & .257$\pm$ .01& .324$\pm$ .02& .126$\pm$ .02 \\
			MMSB   & .512$\pm$ .02& .351$\pm$ .02& \textbf{.725$\pm$ .02} \\
			\hline
		\end{tabular}
	\end{minipage}
	\caption{The experiment result on six datasets. Compared to all other baseline algorithms, we can see that JAG model is consistently the best in most of the metric on all the datasets. }\vspace{-0.3in}
	\label{tab:result}
\end{table*}

\section{Experimental Result}
\label{sec:exp}
In this section, we show the effectiveness of our modeling assumption by comparing JAG with the state-of-the-art community detection methods on the six real-world datasets that are described before.

\subsection{Dataset Preprocessing}
From Table. \ref{tab:datasets}, we can see that the six real-world datasets are all very large scale data.
For experiments we closely replicate \cite{yang2012community}, we sample a large enough graph multiple times to get a large number of small sub-networks with overlapping community structures.  The steps of preprocessing are shown below:
First, we pick a random node $u$ in the given graph $G$ that belongs to at least $k$ communities.
In this paper, we use $k=2$ for preprocessing.
We then take the subnetwork to be the induced subgraph of $G$ consisting of all the nodes that share at least one community membership with the randomly sampled node $u$. Then this subnetwork will be one of the sampled sub-network.
In this experiments, we create $500$ subnetworks for each network.

\subsection{Baseline Methods}
To evaluate the effectiveness of our JAG Model, we choose four baseline algorithms in our experiment: AGM, Link Clustering, Clique Percolation Method, and MMSB.

\textbf{AGM} \cite{yang2012community} is the state-of-the-art community detection algorithm which is our main baseline. We further let AGM use this number of communities by providing the fixed value in the code. This modification ensures fairness of comparisons with our algorithm.

\textbf{LC} (Link clustering) \cite{ahn2010link} is a method that first transfer the adjacent matrix of the graph into a link similarity matrix, and do clustering on the link similarity, then transfer it back to node adjacent metric. For LC, we used the implementation in the Stanford Network Analysis Platform. 

\textbf{CPM} (Clique Percolation Method) \cite{derenyi2005clique} uses Clique percolation to detect network communities. For CPM, we set the communitie size $k$=5 as suggested in \cite{derenyi2005clique}.
We use the Stanford Network Analysis Platform for CPM experiment.

\textbf{MMSB} (Mixed-Membership Stochastic Block Model) \cite{airoldi2008mixed, zhou2015infinite,gopalan2013efficient,ball2011efficient,karrer2011stochastic} combines global parameters that instantiate dense patches of connectivity (blockmodel) with local parameters that instantiate node-specific variability in the connections (mixed membership) to do community detection. We use the ground truth community number as the input of MMSB.  For MMSB we used publicly available 'LDA' R package.

\vspace{-0.1in}
\subsection{Evaluation metrics}
We use three popular evaluation metrics in this paper to evaluate the performance of all the methods: \textbf{F1-score}~\cite{van1979information}, \textbf{Omega Index}~\cite{van1979information}, and \textbf{NMI (Normalized Mutual Information)}~\cite{cover2012elements}.
Due to the page limits, the details of these three metrics can be found in \cite{van1979information,cover2012elements}.

%
%
%
%
All the measures introduced before are in the range of 0 to 1, and a higher value indicates a better community detection result in terms of the ground truth.

\subsection{Result and Discussion}

Following~\cite{yang2012community}, we do all the experiment on the sampled $500$ sub-networks for all the data sets. We measure the detection result on each sub-network by all the three metrics separately. And then we average the result of all the 500 sampled sub-networks.
The result is shown in Table \ref{tab:result}. 

Compared with AGM methods, we can see that JAG model is accurate on all the dataset. This demonstrates that the Jaccard similarity between the community sets of two individuals is the right ``sweet" measure for modeling the edge formation probability. We would like to emphasize that AGM require more community-specific parameter which is hard to estimate, especially for small communities. On the contrary, JAG model only requires one parameter. The better performance of JAG also shows another empirical support of our model.

Compared to all other baseline algorithms, we can see that JAG is consistently achieving the state-of-the-art performance in most of the metric on all the datasets, sometimes significantly outperforming other baselines, including the state-of-the-art AGM. We attribute this to superior and nearly parameter-free modeling which is less prone to over-fitting. Note that JAG with just one parameter can explain several network observation as explained in previews sections.

\section{Conclusion}
\label{sec:con}
We show that Jaccard similarity of the community sets associated with a pair of individuals seems like the ``sweet" measure for modeling edge formation probability. This modeling, unlike popular AGM, does not require to have local parameters for each community, instead, we only need one global hyperparameter to explains most of our observed findings.  We call our parameter-free model Jaccard-based Affiliation Graph (JAG) model.

As a result, we obtain a parameter-free community detection algorithm, which on six real-world datasets, achieves state-of-the-art overlap community detection performances compared with other algorithms, including AGM which requires significantly more parameters.
\vspace{-0.1in}
\section*{Acknowledgments}

This work was supported by National Science Foundation IIS-1652131, RI-1718478, AFOSR-YIP FA9550-18-1-0152, Amazon Research Award, ONR BRC grant on Randomized Numerical Linear Algebra, and a GPU grant from NVIDIA.\vspace{-0.1in}
\bibliographystyle{ieeetran}
\bibliography{jag}

\end{document}